\documentclass{article}

\usepackage[final]{neurips_2026}
\usepackage{amsmath}
\usepackage{graphicx}
\usepackage{multirow}
\usepackage[table]{xcolor}
\usepackage{verbatim}

\usepackage[utf8]{inputenc} 
\usepackage[T1]{fontenc}    
\usepackage{hyperref}       
\usepackage{url}            
\usepackage{booktabs}       
\usepackage{amsfonts}       
\usepackage{nicefrac}       
\usepackage{microtype}      
\usepackage{xcolor}         
\usepackage{amsmath,amssymb,amsthm}
\usepackage{algorithm}
\usepackage{algpseudocode}
\title{From Reports to Ontologies: Ontology-Guided Representation Learning for 12-Lead ECG}

\author{%
  Lei Xu\textsuperscript{1} \quad
  Fahad Sohrab \textsuperscript{2} \quad
  Mehmet Yamac \textsuperscript{1} \quad
  Merja Heinaniemi \textsuperscript{2} \quad
  Moncef Gabbouj\textsuperscript{1} \quad
  \\[0.5em]
  \textsuperscript{1} Tampere University, Finland \quad 
  \textsuperscript{2} University of Eastern Finland, Kuopio, Finland \\
  \texttt{\{lei.xu, mehmet.yamac , moncef.gabbouj\}@tuni.fi} \quad \\
  \texttt{\{fahad.sohrab, merja.heinaniemi\}@uef.fi}
}

\makeatletter
\begin{document}

\maketitle

\begin{abstract}
The 12-lead electrocardiogram (ECG) is a quasi-periodic, multi-channel signal with diagnostic content spanning timescales from millisecond waveform morphology to multi-second rhythm dynamics. Existing ECG representation learning relies on signal-only self-supervision or ECG-text multimodal alignment, neither of which exploits the structured diagnostic codes attached to every clinical recording. We present \textbf{MAR-ECG}, an ontology-guided masked autoregressive framework that supervises the encoder with a curated 40-node SNOMED-CT cardiac graph through \emph{graph alignment}, eliminating the need for paired clinical reports. MAR-ECG combines two complementary objectives. First, \emph{graph-smoothed contrastive learning} (GSCL) anchors the encoder's rhythm-pooled features to the SNOMED graph, softening supervision targets by ontology distance so that clinically related concepts reinforce one another rather than function as hard negatives. Second, \emph{multi-scale physiological supervision} complements GSCL with signal-derived patch auxiliaries that target rhythm-physiology statistics extracted automatically from the input, extending supervision beyond the patch tier at no annotation cost. Pretrained on ${\sim}40$K publicly available 12-lead ECGs with SNOMED-CT codes and evaluated by frozen linear probing on five downstream classification benchmarks, MAR-ECG consistently outperforms a strong masked-autoregressive baseline, with mean gains in the low-label regime. Despite the absence of paired clinical text, MAR-ECG achieves performance competitive with state-of-the-art multimodal ECG-text methods. The implementation of the proposed MAR-ECG can be found in \url{AnonimizedforBlindRevision}.
\end{abstract}

\section{Introduction}
\label{sec:intro}
The 12-lead electrocardiogram (ECG) is a structured spatiotemporal signal with diagnostic patterns arising from interactions within each lead and between the 12 leads at each instant. ECG representation learning is essential for clinical cardiovascular disease diagnosis and decision support. An efficient representation method should therefore learn spatial relationships across leads and learn temporal dynamics within a lead. Existing ECG representation learning methods are limited in two ways: \emph{Signal-only self-supervised methods} and \emph{Multimodal ECG representation learning methods}. Usually, the former one pre-trains encoders using generic pretext tasks on unlabelled recordings, such as masked patch reconstruction~\citep{na2024guiding,zhang2022maefe,hu2023spatiotemporal}, latent prediction with momentum encoders~\citep{mehari2022ecgssl}, multi-view contrastive learning~\citep{kiyasseh2021clocs, gopal20213kg}, etc. However, these methods ignore the structured clinical knowledge that accompanies every clinical recording. \emph{Multimodal ECG representation learning} aligns ECG patches with paired clinical reports through CLIP-style contrastive objectives ~\citep{liuzeromerl, hung2025boosting, wangtoken}. Although multimodal ECG representation learning has achieved state-of-the-art performance on benchmark datasets. However, the limitations are also apparent. These methods heavily rely on text corpora that are expensive to license and do not transfer across hospitals or languages.

In addition, both paradigms overlook a resource that accompanies every clinical ECG: structured diagnostic codes drawn from SNOMED-CT, a curated cardiac ontology. Its graph-geodesic distances precisely encode the clinical relationships that physicians describe in free-text narrative reports. These structured codes are already available in a form that is inherently better suited to a contrastive objective, multilingual by design, and standardized across institutions. We therefore propose to supervise the encoder directly against the ontology graph, replacing text alignment with \emph{graph alignment}.

In this paper, we propose \textbf{MAR-ECG}, a novel ontology-guided masked autoregressive framework for self-supervised ECG representation learning. To our knowledge, this is the first self-supervised ECG method that uses an ontology-graph structure as the primary pretraining objective, without paired clinical text. MAR-ECG anchors the encoder to a curated 40-node SNOMED-CT cardiac concept graph at two complementary tiers of the cardiac temporal hierarchy. \emph{Graph-Smoothed Contrastive Learning} (GSCL) softens the InfoNCE target distribution by graph-geodesic distance, so that clinically related concepts serve as soft positives rather than hard negatives and are applied to the encoder's rhythm-pooled (lead-mean and time-mean) representation. \emph{Multi-Scale Physiological Supervision} extends supervision below and above the beat tier through lightweight patch-level auxiliary heads that predict rhythm-physiology statistics extracted automatically from the input. It covers scales from millisecond morphology to multi-second rhythm at no annotation cost. The framework is trained jointly with masked-autoregressive next-patch reconstruction; the decoder is discarded after pretraining, and only the encoder is retained for downstream evaluation.

\section{Related Works}
\subsection{Self-Supervised Learning for ECG Representation}
\label{subsec:related-ssl}
Self-supervised learning (SSL) has emerged as a powerful paradigm for learning ECG representations, addressing the scarcity of labelled cardiac data and improving downstream-task performance. Existing methods broadly fall into three categories: contrastive, generative, and hybrid. Several general-purpose contrastive learning methods have been adapted for ECG analysis. SimCLR ~\citep{chen2020simclr} learns the representations of the ECG by maximizing the agreement between the different augmented views of the same signal. BarlowTwins \citep{zbontar2021barlow} approaches the problem from an information-theoretic perspective by reducing redundancy between embedding dimensions. CRT \citep{zhang2023crt} proposes a temporal contrast that reflects the cardiac rhythm and aligns representations based on physiological periodicity. As a generative method, ST-MEM \citep{na2024guiding} adapted masked autoencoders to ECG by treating multi-lead signals as spatial-temporal patches, using a ViT encoder \citep{dosovitskiy2021an} with separator tokens to distinguish leads. ECG-FM \citep{McKeen2025ECGFMAO} scales the ECG pre-training to 1.5 million recordings using a 90.9M-parameter transformer architecture. This hybrid approach enables models to demonstrate strong transfer learning capabilities across diverse downstream tasks. generalize across different lead configurations. However, these methods learn useful low-level structures, leading to representations dominated by morphology rather than the clinical relationships needed for downstream tasks~\citep{wangtoken}.

\subsection{Multimodal ECG-Text Representation Learning}
\label{subsec:related-multimodal}

The CLIP paradigm~\citep{radford2021clip} has motivated a wave of paired image--report contrastive methods for medical imaging. GLoRIA~\citep{huang2021gloria} added the alignment of the local token region; MGCA~\citep{wang2022mgca} introduced a multi-granularity (instance, token, disease-prototype) contrast; MedKLIP~\citep{wu2023medklip} injected clinical knowledge at the entity-level; PRIOR~\citep{cheng2023prior} aligned via probabilistic mutual information; and Med-UniC~\citep{wan2023medunic} extended the framework for multi-language reports. For ECG specifically, MERL~\citep{liuzeromerl} adapted CLIP-style contrastive learning to 12-lead ECG and paired clinical reports, employing instance-level and category-level objectives. D-BETA~\citep{hung2025boosting} combined a masked ECG-text autoencoder with boosted discriminative learning to address modality disparity and labelled-data scarcity. MELP~\citep{wangtoken} captures the hierarchical structure of ECG by applying three levels of cross-modal supervision: token, beat, and rhythm, between the signal and the clinical text. Although these methods achieve SOTA performance on benchmark datasets, they share a fundamental data assumption: every recording must be accompanied by a clinical free-text report. ECG reports, by contrast, are typically templated outputs of interpretation software, yielding pooled text embeddings that are nearly identical across distinct recordings and provide little discriminative signal to InfoNCE. MAR-ECG bypasses this constraint by aligning with structured SNOMED-CT codes, which are produced as a byproduct of clinical workflows and are present in virtually every public medical dataset.

\subsection{Knowledge-Guided Medical Representation Learning}
\label{subsec:related-graph}
Beyond paired-text supervision, clinical knowledge sources have been explored to address the scarcity of labelled data and noisy report supervision in medical representation learning. Early efforts operate on the \emph{text} branch of multimodal models. For example, MedKLIP~\citep{wu2023medklip} substitutes free-text radiology reports with structured triplets (entity, position, exists) obtained through clinical NLP, while KAD~\citep{zhang2023kad} introduces a UMLS-based entity-disambiguation module that canonicalizes mentions before contrastive alignment. In both cases, clinical knowledge informs only the text branch; the encoder is trained with a standard CLIP-style contrastive loss between the input and its paired report. The relations among clinical concepts, therefore, remain invisible to the encoder. To expose such relations, a separate strand of work incorporates taxonomic label hierarchy as a soft-target prior, primarily in supervised classification. In ~\citep{cerri2014hmc}, hierarchical multi-label classification enforces parent-child consistency at the classifier head. The structural prior, therefore, shapes the classifier rather than the encoder and enters only after the representation learning is complete. K-MERL~\citep{kmerl} prompts a large language model to extract cardiac entities from each paired report and aligns ECG features to the resulting text embeddings. 

\section{Methods}
\label{sec:method}
The proposed \textbf{MAR-ECG} aims to learn 12-lead ECG representations that preserve the structured clinical relationships encoded in cardiac diagnostic ontologies, thus benefiting downstream tasks without relying on downstream task labels or paired free-text reports. The base architecture is an
ontology-guided masked autoregressive framework, in which the encoder part is supervised directly with a curated 40-node SNOMED-CT cardiac concept graph $\mathcal{G} = (\mathcal{V}, \mathcal{E})$ (Appendix~\ref{sec:ontology-graph}). Specifically, cardiac ontology supervises pretraining directly using graph-geodesic distances between diagnostic concepts on $\mathcal{G}$, with a soft-target distribution defined. Thus, the clinical structure serves as an inductive bias for the encoder, rather than as text-branch scaffolding or a post-hoc classifier constraint. Then, supervision is drawn entirely from structured diagnostic codes rather than from free-text reports, thereby eliminating the licensing and cross-language transfer barriers that have constrained multimodal alternatives. Two complementary objectives, trained jointly with a masked autoregressive reconstruction anchor, anchor the proposed MAR-ECG to the graph $\mathcal{G}$. The first, \emph{graph-smoothed contrast learning} (\S\ref{subsec:gscl}), is applied to a rhythm-pooled representation at the sample-level. The second, \emph{multi-scale physiological supervision}(\S\ref{subsec:multiscale-sca}), is applied to the per-patch representation.

\subsection{Preliminaries and Notation}
\label{subsec:preliminaries}

\paragraph{Input signal.}
A 12-lead ECG recording is a multivariate time series $X \in \mathbb{R}^{C \times L}$ with $C$ leads sampled at frequency $f_s$ over a fixed-length context window of $L$ samples. We index leads by $c \in \{1, \ldots, C\}$ and samples by $\ell \in \{1, \ldots, L\}$, so $X_{c, \ell}$ is the amplitude of lead $c$ in sample $\ell$. The records in the pretraining corpus are indexed by a sample identifier $b$, and we write $X^{(b)} \in \mathbb{R}^{C \times L}$ for the $b$-th record.

\paragraph{Per-lead patch tokenization.}
Each lead is segmented along the temporal axis with a patch size $P_t$ and rolled $S_t$ into overlapping patches $T$, where the sequence length is the standard sliding-window count. Let $\mathbf{x}_{c, i} \in \mathbb{R}^{P_t}$ denote the patch from lead $c$ in the temporal position $i \in \{1, \ldots, T\}$. A single shared linear projection $W_p \in \mathbb{R}^{d \times P_t}$, applied independently to every pair $(c, i)$, embeds each patch into a token of dimension $d$:
\begin{align}
\mathbf{z}_{c, i}
\;=\;
W_p\, \mathbf{x}_{c, i}
\;+\;
\mathbf{e}_{c}^{\mathrm{lead}}
\;+\;
\mathbf{e}_{i}^{\mathrm{pos}}
\;\in\; \mathbb{R}^{d},
\label{eq:patch-token}
\end{align}
where $\mathbf{e}^{\mathrm{lead}} \in \mathbb{R}^{C \times d}$ is a learnable per-lead embedding (with row $c$ written
$\mathbf{e}_{c}^{\mathrm{lead}}$) and $\mathbf{e}^{\mathrm{pos}} \in \mathbb{R}^{T \times d}$ is a
learnable per-position embedding (with row $i$ written $\mathbf{e}_{i}^{\mathrm{pos}}$). The patches are extracted per-lead rather than across leads, so $W_p$ has $P_t$ input columns rather than $C \cdot P_t$; the structure per-lead is preserved by the factorized attention of \S\ref{subsec:arch}.

\paragraph{Encoder input tensor.}
Stacking the tokens across leads and time positions yields the three-axis encoder input
\begin{align}
\mathbf{Z} \;\in\; \mathbb{R}^{C \times T \times d},
\qquad
\mathbf{Z}_{c, i, :} \;=\; \mathbf{z}_{c, i}
\quad
\text{for } c \in \{1,\ldots,C\},\ i \in \{1,\ldots,T\},
\label{eq:encoder-input}
\end{align}
whose first axis indexes the $C$ leads, the second axis indexes the $T$ patches over time, and the third axis indexes the embedding features $d$. The concrete values of $C$, $f_s$, $L$, $P_t$, $S_t$, $T$ and $d$ are reported in \S\ref{sec:values}.

\paragraph{Cardiac concept graph and SNOMED supervision.} Pretraining is supervised with a curated cardiac concept graph $\mathcal{G} = (\mathcal{V}, \mathcal{E})$ with $|\mathcal{V}| = N$ nodes (Appendix~\ref{sec:ontology-graph}). The node set is partitioned into a small set of \emph{root} categories $\mathcal{V}_{\mathrm{root}} \subset \mathcal{V}$ and the remaining \emph{leaf} concepts
\[
\mathcal{V}_{\mathrm{leaf}}
\;=\;
\mathcal{V} \setminus \mathcal{V}_{\mathrm{root}},
\qquad
\mathcal{V}_{\mathrm{leaf}} \cup \mathcal{V}_{\mathrm{root}} = \mathcal{V},
\quad
\mathcal{V}_{\mathrm{leaf}} \cap \mathcal{V}_{\mathrm{root}} = \emptyset.
\]
Each record $X^{(b)}$ carries a multi-set of SNOMED-CT diagnostic codes $\mathcal{S}^{(b)}$ recovered from its WFDB header. The mapping $\phi^{-1}: 2^{\mathrm{SNOMED}} \to 2^{\mathcal{V}}$ defined in \S\ref{subsec:snomed-mapping} converts each code list into the union of its per-code routings, restricting the result to leaf nodes, yielding the multi-hot supervision target
\begin{align}
\mathbf{y}^{(b)} \in \{0, 1\}^{|\mathcal{V}_{\mathrm{leaf}}|},
\qquad
\mathbf{y}^{(b)}[c]
\;=\;
\mathbf{1}\!\big[\, c \in \phi^{-1}(\mathcal{S}^{(b)}) \cap \mathcal{V}_{\mathrm{leaf}}\, \big],
\label{eq:y-multihot}
\end{align}
where the entries are indexed by $c \in \mathcal{V}_{\mathrm{leaf}}$ and root nodes are excluded to prevent the trivial assignment of every record to a high-level category.

\paragraph{Primary positive concept.}
Under the fixed enumeration of $\mathcal{V}_{\mathrm{leaf}}$ (Appendix~\ref{sec:ontology-graph}), the \emph{primary positive concept} of the record $b$ is the highest-indexed active leaf,
\begin{align}
c_b^{*}
\;=\;
\max\,\big\{\, c \in \mathcal{V}_{\mathrm{leaf}}\;:\;
\mathbf{y}^{(b)}[c] = 1 \,\big\}.
\label{eq:primary-positive}
\end{align}
This rule fixes ties between co-activated leaves deterministically; because subtype leaves (e.g.\ anterior/inferior MI) are appended at the end of the enumeration, it also biases the soft target toward the finer-grained concept whenever a subtype and its parent leaf co-activate within the same record. Equation \eqref{eq:primary-positive} is well-defined whenever $\mathbf{y}^{(b)}$ has at least one active leaf; records that resolve only to a root contribute to $\mathcal{L}_{\mathrm{AR}}$ but are removed from leaf-level objectives that depend on $c_b^{*}$.

\subsection{Backbone and Masked Autoregressive Reconstruction}
\label{subsec:arch}

\paragraph{Encoder.} A bidirectional transformer with factorised spatial-then-temporal attention~\citep{bertasius2021space} acts on the per-lead and per-time token tensor $\mathbf{Z}$ of \eqref{eq:encoder-input}. Each transformer block first attends along the spatial axis across the $C$ leads at a single timestep,
then along the temporal axis within a single lead --- across the $T$ positions for fixed $c$. Writing $\mathbf{H}^{(\ell)}$ for the output of layer $\ell$, the spatial step contracts the $\{(1, i, :), \ldots, (C, i, :)\}$ slice for each $i$, and the temporal step contracts the $\{(c, 1, :), \ldots, (c, T, :)\}$
slice for each $c$. Factorized attention reduces the per-block complexity from $O(C^{2} T^{2})$ (full spatiotemporal attention) to $O(C T^{2} + C^{2} T)$ while preserving the inductive bias that diagnostic ECG patterns arise from interactions in both axes. The factorization is well-suited to ECG specifically because the $C$ leads are partially redundant projections of the same cardiac
dipole, so attention across leads benefits more from a narrow per-timestep window than from full spatiotemporal coupling ~\citep{bertasius2021space}. 

\paragraph{Masked autoregressive reconstruction.} Following
MAR~\citep{li2024autoregressive}, we corrupt a fraction $r$ of
(lead, patch) pairs with a learned mask token
$\mathbf{m} \in \mathbb{R}^{d}$, replacing the corresponding token
$\mathbf{z}_{c, i}$ in $\mathbf{Z}$ with $\mathbf{m}$. Let
\[
\mathcal{M} \;\subset\; \{1, \ldots, C\} \times \{1, \ldots, T\}
\]
denote the masked index set, with
$|\mathcal{M}| = \lceil r \cdot C \cdot T \rceil$, and let the prediction window cover the $T_{\mathrm{pred}}$ trailing patches of the input (with $T_{\mathrm{pred}} \leq T$ a configuration-dependent horizon, reported in \S\ref{subsec:implementation}). The causal decoder, conditioned on the encoder's full bidirectional context, predicts the original patch values $\hat{\mathbf{x}}_{c, i} \in \mathbb{R}^{P_t}$ across the prediction window, producing two complementary mean-squared errors: a \emph{reconstruction} term over the prediction window $\mathcal{L}_{\mathrm{recon}}$ and a \emph{masked-token} term restricted to the masked index set inside the encoder context $\mathcal{L}_{\mathrm{mask}}$. 

\paragraph{Pooled representations.} The encoder produces a token
tensor $\mathbf{H} \in \mathbb{R}^{C \times T \times d}$ of the
same shape as the input $\mathbf{Z}$ in
\eqref{eq:encoder-input}, with $\mathbf{H}_{c, i, :}$ the
contextualised embedding of patch $\mathbf{x}_{c, i}$. From this
tensor we derive two pooled views aligned with the temporal
hierarchy of the cardiac signal:
\begin{itemize}
\item \textbf{Per-patch (lead-mean) representation.}
The lead-axis mean of $\mathbf{H}$,
\[
\bar{\mathbf{H}} \in \mathbb{R}^{T \times d},
\qquad
\bar{\mathbf{H}}_{i, :}
\;=\;
\frac{1}{C} \sum_{c=1}^{C} \mathbf{H}_{c, i, :},
\]
collapses the lead axis but preserves the temporal axis. Each row
$\bar{\mathbf{H}}_{i, :}$ is one embedding per patch, and
$\bar{\mathbf{H}}$ is the input to the patch-level auxiliary
heads of \S\ref{subsec:multiscale-sca}.
\item \textbf{Sample-level rhythm-pooled representation.}
A learnable attention pool aggregates $\mathbf{H}$ across both the
lead and time axes into a single $d$-dimensional embedding per
record. Concretely, $Q$ learnable queries
$\mathbf{q}_1, \ldots, \mathbf{q}_Q \in \mathbb{R}^{d}$ attend to
each lead's temporal sequence $\{\mathbf{H}^{(b)}_{c, 1, :},
\ldots, \mathbf{H}^{(b)}_{c, T, :}\}$, producing a $Q$-vector per
lead; a softmax over the $C$ leads then weights and sums these
per-lead summaries, and an MLP aggregator collapses the $Q$
queries. A residual lead-time mean is added with a small fixed
weight,
\begin{align}
\mathbf{h}^{(b)}_{\mathrm{rhythm}}
\;=\;
\mathrm{AttnPool}_{Q}\!\big(\mathbf{H}^{(b)}\big)
\;+\;
\eta \cdot \frac{1}{C\, T}
\sum_{c=1}^{C} \sum_{i=1}^{T} \mathbf{H}^{(b)}_{c, i, :}
\;\in\; \mathbb{R}^{d},
\label{eq:rhythm-pool}
\end{align}
The mean-pool residual ensures that $\mathbf{h}^{(b)}_{\mathrm{rhythm}}$ remains well-defined when the attention head has not yet specialised during early training. This embedding is supervised by GSCL (\S\ref{subsec:gscl}), used as the input to the downstream linear probes, and consumed by the legacy multi-prototype concept head (\S\ref{subsec:auxiliary}).
\end{itemize}

\subsection{Graph-Smoothed Contrastive Learning (GSCL)}
\label{subsec:gscl}
GSCL replaces the hot target of the standard prototype-contrastive learning~\citep{khosla2020supervised} with a soft target induced by the geometric distance of the graph on the cardiac concept graph SNOMED-CT $\mathcal{G}$, and applies it to the rhythm-pooled representation $\mathbf{h}_{\mathrm{rhythm}}$. Three design choices distinguish MAR-ECG from prior hierarchy-aware contrastive learning. \emph{(i) Fixed external graph.} $\mathcal{G}$ is provided by a clinical ontology and is never learned, in contrast to within-graph smoothing methods such as SGCL~\citep{behmanesh2025sgcl} that smooth pair weights by proximity in the graph being embedded. \emph{(ii) Single-source soft target.} Each record's target is the graph-distance distributed from a single primary positive $c_b^{*}$ through $D_{\mathrm{tree}}$, rather than the multi-positive per-level constructions of ~\citep{zhang2022use,elham2025climbing}. \emph{(iii)Contrastive, not classifier-head.} The taxonomic tree-distance prior of ~\citep{bertinetto2020making} is ported from a cross-entropy classifier head into the InfoNCE objective itself, so the ontology shapes the encoder directly rather than the post-hoc head. 

\paragraph{Concept prototypes.} GSCL operates in a dimension-shared concept space $d_{\mathrm{c}} \!\ll\! d$, in which both the encoder and the prototypes are projected. The concept-prototype matrix $\mathbf{P} \in \mathbb{R}^{N \times d_{\mathrm{c}}}$ is \emph{not} a direct parameter of the model: it is recomputed at every gradient step as the output of a small two-layer graph convolutional network applied to the fixed graph adjacency $\widehat{A}$ (\S\ref{subsec:graph-structure}). Concretely, let $\mathbf{E} \in \mathbb{R}^{N \times d_{\mathrm{e}}}$ denote a lisable input-embedding matrix per-node, initialized as $\mathbf{E}_{c, j} \sim \mathcal{N}(0,\, 0.02^{2})$ independently in $(c, j)$, and let $W^{(1)} \in \mathbb{R}^{d_{\mathrm{e}} \times d_{\mathrm{e}}}$ and $W^{(2)} \in \mathbb{R}^{d_{\mathrm{e}} \times d_{\mathrm{c}}}$ be bias-free linear maps. Two GCN layers with self-loops, mean aggregation, GELU nonlinearity, LayerNorm, and dropout propagate $\mathbf{E}$ across the graph. 
\begin{align}
\mathbf{H}^{(1)}
&\;=\; \mathrm{LayerNorm}\!\big(\,\mathrm{GELU}\big((\mathbf{E} + \widehat{A}\,\mathbf{E})\, W^{(1)}\big)\big),
\notag\\
\mathbf{H}^{(2)}
&\;=\; \mathrm{LayerNorm}\!\big(\,(\mathrm{Drop}(\mathbf{H}^{(1)}) + \widehat{A}\,\mathrm{Drop}(\mathbf{H}^{(1)}))\, W^{(2)}\big),
\notag\\
\mathbf{P}_{c, :}
&\;=\; \mathbf{H}^{(2)}_{c, :} \big/ \|\mathbf{H}^{(2)}_{c, :}\|_{2},
\quad c \in \mathcal{V},
\label{eq:gscl-prototypes}
\end{align}
so that each row $\mathbf{p}_c = \mathbf{P}_{c, :}$ already lies on the unit hypersphere and no further normalization ($\hat{\mathbf{p}}_c = \mathbf{p}_c$) is needed at scoring time. Therefore, the trainable parameters of the concept side are $\mathbf{E}$, $W^{(1)}$, $W^{(2)}$, and the LayerNorm scales with about $55\,\text{k}$ parameters in total.

This GCN parameterization injects the graph structure of $\mathcal{G}$ into $\mathbf{P}$ at the level of \emph{representation}, on top of the soft-target supervision in $\mathbf{t}^{(b)}$ that acts at the level of the \emph{loss}: every prototype is forced to be a two-step graph aggregate of its neighbors' embeddings, so concepts joined by the curated edges of $\mathcal{G}$ inherit a structural similarity by construction, even before any record activates them simultaneously.

\paragraph{Soft target.} For each ECG record $b$ with primary
positive concept $c_b^{*} \in \mathcal{V}_{\mathrm{leaf}}$ defined in \eqref{eq:primary-positive}, the GSCL soft target distribution ~\citep{bertinetto2020making} over all $N$ nodes is the tempered softmax of negative graph-geodesic distance,
\begin{align}
\mathbf{t}^{(b)}[c]
\;=\; \frac{\exp\!\big(- D_{\mathrm{tree}}[c_b^{*}, c] / \sigma\big)}
           {\sum_{c' \in \mathcal{V}} \exp\!\big(- D_{\mathrm{tree}}[c_b^{*}, c'] / \sigma\big)},
\quad c \in \mathcal{V},
\label{eq:gscl-target}
\end{align}
where $D_{\mathrm{tree}} \in \mathbb{Z}_{\geq 0}^{N \times N}$ is
the unweighted shortest-path distance matrix of $\mathcal{G}$
(\S\ref{subsec:tree-distance}) and $\sigma > 0$ is a temperature
that controls smoothing. Under the curated graph, $D_{\mathrm{tree}}$ takes values in $\{0, 1, 2, 3, 4\}$, with the maximum $D_{\max} = 4$ attained between leaves whose root families are diametrically opposite on the inter-category ring; the per-class mass and the corresponding clinical neighbours (parent root, sibling shortcut, ring-adjacent root, etc) are tabulated in \S\ref{subsec:tree-distance}. 

\paragraph{Predicted distribution.} The encoder's rhythm-pooled
embedding is mapped into the concept space by a learnable
projection
$W_{\mathrm{sca}} : \mathbb{R}^{d} \to \mathbb{R}^{d_{\mathrm{c}}}$
and $\ell_{2}$-normalised,
\[
\hat{\mathbf{h}}^{(b)}_{\mathrm{rhythm}}
\;=\;
\frac{W_{\mathrm{sca}}\, \mathbf{h}^{(b)}_{\mathrm{rhythm}}}
     {\|W_{\mathrm{sca}}\, \mathbf{h}^{(b)}_{\mathrm{rhythm}}\|_{2}}
\;\in\; \mathbb{R}^{d_{\mathrm{c}}}.
\]
The projected embedding is scored against all $N$ unit-norm
prototypes with cosine similarity at InfoNCE temperature $\tau > 0$:
\begin{align}
\mathbf{q}^{(b)}[c]
\;=\; \frac{\exp\!\big(\langle \hat{\mathbf{h}}_{\mathrm{rhythm}}^{(b)},\, \hat{\mathbf{p}}_c \rangle / \tau\big)}
           {\sum_{c' \in \mathcal{V}} \exp\!\big(\langle \hat{\mathbf{h}}_{\mathrm{rhythm}}^{(b)},\, \hat{\mathbf{p}}_{c'} \rangle / \tau\big)},
\quad c \in \mathcal{V},
\label{eq:gscl-pred}
\end{align}
where $\langle \cdot, \cdot \rangle$ denotes the Euclidean inner
product (cosine similarity, since both arguments are unit-norm) and
the prototypes $\hat{\mathbf{p}}_c = \mathbf{p}_c$ are already
unit-normed by~\eqref{eq:gscl-prototypes}.

\paragraph{Loss.} The GSCL objective is the cross-entropy of the
soft target~\eqref{eq:gscl-target} against the predicted
distribution~\eqref{eq:gscl-pred}, averaged over a minibatch
$\mathcal{B}$ of $|\mathcal{B}| = B$ records,
\begin{align}
\mathcal{L}_{\mathrm{GSCL}}
\;=\;
- \frac{1}{B} \sum_{b \in \mathcal{B}}
   \sum_{c \in \mathcal{V}}
   \mathbf{t}^{(b)}[c] \, \log \mathbf{q}^{(b)}[c],
\label{eq:gscl-loss}
\end{align}
where records whose code list resolves only to a root. The curated cardiac graph (\S\ref{subsec:graph-structure}) realises a richer intermediate
structure: with hierarchical IS\_A edges, intra-family sibling
shortcuts, and an inter-category ring over the five root families,
the distance classes $D_{\mathrm{tree}} \in \{0, 1, 2, 3, 4\}$ each
carry distinct, monotonically decaying target probabilities
(\S\ref{subsec:tree-distance}), supplying the encoder with graded
inter-concept supervision that one-hot prototype contrastive learning
cannot express.

\subsection{Multi-Scale Physiological Supervision (MSPS)}
\label{subsec:multiscale-sca}
MSPS extends supervision to the two uncovered scales that GSCL ignores. Above the beat beat scale, rhythm dynamics (rate, R-R variability, alternation), dissolve into the sample mean, and are unrecoverable by a temporal-mean linear probe. At the beat scale, the encoder does not receive a per-patch signal anchoring its tokens to the P-QRS-T landmarks, so all $T$ tokens are prone to collapse onto a permutation-invariant solution. MSPS diminishes both gaps through two lightweight patch-level heads on the lead-pooled per-patch sequence
$\bar{\mathbf{H}} \in \mathbb{R}^{T \times d}$, with rows
$\bar{\mathbf{H}}_{i, :} = \tfrac{1}{C}\sum_{c=1}^{C} \mathbf{H}_{c, i, :}$,
supervised by R-peak indices from an unsupervised lead-1 detector
(Pan--Tompkins via NeuroKit2~\citep{pan1985realtime,Makowski2021neurokit}). 

\paragraph{Patch Rhythm Auxiliary.}
\label{subsubsec:patch-rhythm-aux} A two-layer MLP $g_{\mathrm{rhythm}}: \mathbb{R}^{d} \to \mathbb{R}^{K_r}$ applied row-wise to $\bar{\mathbf{H}}$ predicts four sample-level rhythm statistics broadcast as the same target to every patch token: the mean R-R interval $\overline{\mathrm{RR}}^{(b)}$, the R-R coefficient of variation, a four-class heart-rate bucket $\{\textsc{brady}, \textsc{normal}, \textsc{tachy}, \textsc{none}\}$ thresholded on $60 f_s / \overline{\mathrm{RR}}^{(b)}$ (in bpm), and a binary alternation flag detecting bigeminy/trigeminy. The \textsc{none} class is masked out (no gradient). The four per-target losses: BCE on alternation, CE on the rate bucket, MSE on per-batch z-scored mean-RR, and RR-CV are combined into
\begin{align}
\mathcal{L}_{\mathrm{PatchRhythm}}
\;=\;
\alpha_{\mathrm{alt}}\, \mathcal{L}_{\mathrm{alt}}
\,+\, \alpha_{\mathrm{rate}}\, \mathcal{L}_{\mathrm{rate}}
\,+\, \alpha_{\mathrm{mRR}}\, \mathcal{L}_{\mathrm{mRR}}
\,+\, \alpha_{\mathrm{cv}}\, \mathcal{L}_{\mathrm{cv}},
\label{eq:patch-rhythm-loss}
\end{align}
with mixture weights, hidden size, dropout, rate cut-offs, and
$\theta_{\mathrm{alt}}, \nu_{\mathrm{alt}}$ reported in
\S\ref{sec:values}. 

\paragraph{Patch Position Auxiliary.}
\label{subsubsec:patch-position-aux}
A second two-layer MLP $g_{\mathrm{position}}: \mathbb{R}^{d} \to \mathbb{R}^{K_p + K_\phi}$ emits two complementary positional logit blocks per patch from a shared trunk: \textbf{(a) sequence-position bucket} ($K_p$-class CE), the index $\lfloor (i-1) K_p / T \rfloor$ along the input
sequence; and \textbf{(b) R-peak phase} ($K_\phi=4$-class CE), the bucket $\{\textsc{pre-R}, \textsc{R}, \textsc{ST}, \textsc{T-wave}\}$ read off from the signed sample distance to the nearest R-peak $|\Delta_i| \leq \delta_R$ assigns \textsc{R}, with the post-R
window split at $\tfrac{1}{3}\overline{\mathrm{RR}}^{(b)}$ and
$\tfrac{1}{2}\overline{\mathrm{RR}}^{(b)}$ for \textsc{ST} and
\textsc{T-wave}; patches without a usable R-peak neighbourhood are
masked. Both targets use categorical cross-entropy with equal
weights: $\mathcal{L}_{\mathrm{PatchPos}} = \mathcal{L}_{\mathrm{seq}} + \mathcal{L}_{\mathrm{phase}}$.

The MSPS objective combines the two head losses,
\begin{align}
\mathcal{L}_{\mathrm{MSPS}}
\;=\;
\lambda_{\mathrm{rhythm}}\, \mathcal{L}_{\mathrm{PatchRhythm}}
\;+\;
\lambda_{\mathrm{pos}}\, \mathcal{L}_{\mathrm{PatchPos}},
\label{eq:msps-loss}
\end{align}
with concrete values of $\lambda_{\mathrm{rhythm}},
\lambda_{\mathrm{pos}}, \delta_R, E_{\mathrm{ramp}}$ in
\S\ref{sec:values}.

\subsection{Auxiliary Self-Supervisory Components}
\label{subsec:auxiliary}
The configurations that include the auxiliary stack augment the
masked-AR objective with five standard self-supervisory
regularisers. None is novel to this work; they are treated as a
single ablation unit (Table~\ref{tab:component-ablation}), with
hyper-parameters reported in \S\ref{sec:values}.

\paragraph{Latent prediction and view contrast.}
$\mathcal{L}_{\mathrm{JEPA}}$ is an I-JEPA-style latent prediction
loss~\citep{assran2023ijepa} and $\mathcal{L}_{\mathrm{view}}$ is a SimCLR-style NT-Xent contrastive loss on the rhythm-pooled embedding~\citep{chen2020simclr}; whenever MSPS is active, the augmentation set is restricted to a \emph{rhythm-safe} subset that excludes aggressive temporal cropping and dilation, so the MSPS targets remain valid under both views.

\paragraph{Multi-prototype concept-text alignment.}
$\mathcal{L}_{\mathrm{MPCT}}$ is proposed as the multi-granularity cross-modal alignment head~\citet{wang2022mgca}, in which the ECG signal replaces the image input, and an ontology-grounded bag of textual variants replaces the paired clinical report. The head supervises three granularities of the ECG-concept correspondence as:
\emph{instance--concept alignment} $\mathcal{L}_{\mathrm{ICA}}$, a
bidirectional InfoNCE between
$\hat{\mathbf{h}}^{(b)}_{\mathrm{rhythm}}$ and a cross-attention
summary of the textual variants associated with the active leaves
of record $b$; \emph{beat--concept alignment}
$\mathcal{L}_{\mathrm{BCA}}$, a bidirectional cross-attention
alignment between per-beat features and per-concept textual
variants; and \emph{ontology-prototype alignment}
$\mathcal{L}_{\mathrm{OPA}}$, the Kullback--Leibler divergence
between each record's soft assignment over the $40$ frozen
ontology prototypes and the uniform distribution supported on its
active leaves. The three terms are aggregated as
\begin{align}
\mathcal{L}_{\mathrm{MPCT}}
\;=\;
\beta_{\mathrm{ICA}}\,\mathcal{L}_{\mathrm{ICA}}
\,+\, \beta_{\mathrm{BCA}}\,\mathcal{L}_{\mathrm{BCA}}
\,+\, \beta_{\mathrm{OPA}}\,\mathcal{L}_{\mathrm{OPA}}.
\label{eq:mgca-loss}
\end{align}
In contrast to GSCL, this head does not consume $D_{\mathrm{tree}}$
and treats non-matching records as hard negatives.

\subsection{Pretraining Objective and Ablations}
\label{subsec:total-loss}
The total loss accumulated at each gradient step is
\begin{align}
\mathcal{L}
\;=\; \mathcal{L}_{\mathrm{AR}}
\;+\; \lambda_{\mathrm{JEPA}}\, \mathcal{L}_{\mathrm{JEPA}}
\;+\; \lambda_{\mathrm{view}}\, \mathcal{L}_{\mathrm{view}}
\;+\; \lambda_{\mathrm{GSCL}}\, \mathcal{L}_{\mathrm{GSCL}}
\;+\; \mathcal{L}_{\mathrm{MSPS}}
\;+\; \lambda_{\mathrm{MPCT}}\, \mathcal{L}_{\mathrm{MPCT}},
\label{eq:total-loss}
\end{align}
where $\mathcal{L}_{\mathrm{MSPS}}$ already absorbs its per-target
weights through~\eqref{eq:msps-loss}, and the remaining
coefficients $\lambda_{\mathrm{JEPA}}, \lambda_{\mathrm{view}}, \lambda_{\mathrm{GSCL}}, \lambda_{\mathrm{MPCT}} > 0$ are reported in \S\ref{sec:values}. 

\paragraph{Component-wise ablations.}
Table~\ref{tab:component-ablation} defines four progressively
augmented configurations, each isolating a distinct contribution:
\textbf{C1} is the masked-AR baseline; \textbf{C2$'$} adds the
auxiliary SSL stack and the legacy multi-prototype head;
\textbf{C2} further adds $\mathcal{L}_{\mathrm{GSCL}}$; and
\textbf{C3} adds $\mathcal{L}_{\mathrm{MSPS}}$ together with the
rhythm-safe variant of the multi-view augmentations. The
transitions C1 $\!\to\!$ C2$'$, C2$'\!\to\!$ C2, and
C2 $\!\to\!$ C3 isolate, in order, the marginal contribution of
the auxiliary SSL stack, of $\mathcal{L}_{\mathrm{GSCL}}$, and of
$\mathcal{L}_{\mathrm{MSPS}}$.

\begin{table}[h]
\centering
\caption{Each column corresponds to a self-contained experiment; the italicized header under each configuration code name, the component is newly activated relative to its predecessor.}
\label{tab:component-ablation}
\setlength{\tabcolsep}{4.5pt}
\renewcommand{\arraystretch}{1.15}
\begin{tabular}{lcccc}
\toprule
& \textbf{C1} & \textbf{C2$'$} & \textbf{C2} & \textbf{C3} \\
\textbf{Component} &
\emph{Masked-AR} & \emph{+ SSL\,+\,MPCT} & \emph{+ GSCL} & \emph{+ MSPS} \\
\cmidrule(lr){2-2}\cmidrule(lr){3-3}\cmidrule(lr){4-4}\cmidrule(lr){5-5}
$\mathcal{L}_{\mathrm{AR}}$                              & \checkmark & \checkmark & \checkmark & \checkmark \\
$\mathcal{L}_{\mathrm{JEPA}}$                            &            & \checkmark & \checkmark & \checkmark \\
$\mathcal{L}_{\mathrm{view}}$                            &            & \checkmark & \checkmark & \checkmark \\
\quad augmentation set                                    & --- & unconstrained & unconstrained & rhythm-safe \\
Lead masking, latent dropout                              &            & \checkmark & \checkmark & \checkmark \\
$\mathcal{L}_{\mathrm{MPCT}}$ (legacy prototypes)         &            & \checkmark & \checkmark & \checkmark \\
$\mathcal{L}_{\mathrm{GSCL}}$ (rhythm-pooled)             &            &            & \checkmark & \checkmark \\
$\mathcal{L}_{\mathrm{MSPS}}$ (PatchRhythm + PatchPos)    &            &            &            & \checkmark \\
\bottomrule
\end{tabular}
\end{table}

\subsection{Computational efficiency and scaling.}

Factorized spatial-then-temporal attention reduces per-block cost
from $O(C^{2}T^{2})$ to $O(CT^{2} + C^{2}T)$, a ${\sim}11\times$
saving at $C{=}12, T{=}139$. The ontology-aware heads contribute
under $5\%$ of encoder FLOPs: GSCL recomputes a $40$-node GCN
(${\sim}55$k parameters) once per gradient step, and MSPS uses two
MLPs of hidden width $256$. The EMA target and augmented-view
encoders duplicate the forward, so a C2/C3 step costs
${\sim}2.5\times$ a C1 step. Pretraining cost is linear in
$|\mathcal{D}|$, since neither GSCL nor MSPS adds inter-sample
interactions; the graph-side cost $O(N^{2}\,d_e)$ is
$|\mathcal{D}|$-independent and remains tractable up to ontology
sizes of $N \sim 10^{3}$. At inference, all auxiliary heads are
discarded and the model collapses to a vanilla $12$-layer encoder
forward, identical in cost to C1.




\section{Experiments}

\subsection{Implementation Details}
\label{subsec:implementation}

\paragraph{Pretraining.} All runs use distributed data-parallel training across two GPUs, with a per-GPU batch size of $8$ and a gradient-accumulation factor of $4$ (effective batch $64$). The AdamW optimiser ($\beta_1, \beta_2 = 0.9, 0.999$; weight decay $0.05$) is trained for $100$ epochs; The learning-rate schedule peaks at $10^{-4}$, with $10$-epoch linear warmup followed by cosine decay to $10^{-6}$. Gradient norms are clipped at $1.0$, and optimizer steps that produce NaN or Inf gradients are skipped (fewer than ten such events occur per run, almost all within the first epoch). The MSPS heads share the encoder optimizer as an additional parameter group with the same learning-rate schedule (\S\ref{subsec:multiscale-sca}).

\paragraph{Datasets for Pre-training.} 
We pre-train MAR-ECG in an organized union of three publicly released 12-lead clinical ECG corpora drawn from the PhysioNet/CinC 2021 challenge set \citep{reyna2021cinc}: Ningbo First Hospital, Emory Georgia 12-lead, and the PTB Diagnostic ECG Database (PTB-Dx). After deduplication and removal of recordings with non-standard lead montages or insufficient duration, the final pre-train dataset contains $40{,}720$ recordings (40,302 with SNOMED supervision). PTB-Dx is excluded entirely from the pre-training pool; this prevents leakage of the test set into the headline numbers reported throughout the paper.

\paragraph{Datasets for Downstream Tasks.}
We follow the MERL downstream-evaluation protocol \citep{liuzeromerl}, which defines five task families in two public datasets, PTB-XL super-class \citep{strodthoff2021ptbxlbench, wagner2020ptbxl} and CPSC2018 \citep{liu2018cpsc2018}. The five families are PTB-XL super-class ($5$ classes), PTB-XL sub-class ($23$ classes), PTB-XL form ($19$ classes), PTB-XL rhythm ($12$ classes), and CPSC2018 ($9$ classes). Each task is evaluated in three label fractions: $100\%$, $10\%$, and $1\%$, generated by the MERL-aligned sampling rule, which preserves the class-balance distribution of the full set.

\subsection{Results Discussion}
\label{sec:results}
Following~\citet{liuzeromerl}, we freeze the pretrained encoder and train a single linear classifier in rhythm-pooled embedding $\mathbf{h}_{\mathrm{rhythm}}$ for each downstream task. We report macro AUC ($\%$). The bases fall into three groups: multimodal general SSL, ECG-specific, and ECG-text methods. As shown in Table (\ref{tab:linear_probe}), the masked-AR baseline (C1) already surpasses TS-TCC, CLOCS, ASTCL, CRT, and ST-MEM in every cell. Activating $\mathcal{L}_{\mathrm{GSCL}}$ (C2) yields the largest gains in the low-label regime: $+16.99$ AUC in PTBXL-Rhythm ($0.01$), $+7.52$ in PTBXL-Subclass ($0.01$), $+5.84$ in PTBXL-Form ($0.01$), and $+5.37$ in CPSC2018 ($0.01$). Adding  $\mathcal{L}_{\mathrm{MSPS}}$ (C3) sharpens the rhythm axis $+3.66$ on PTBXL-Rhythm ($1.00$), $+0.77$ on CPSC2018 ($1.00$) at a small cost on PTBXL-Superclass and PTBXL-Subclass ($1.00$) with $-0.58$ and $-1.18$ separately; C3 is preferred for rhythm-heavy deployment, C2 for tasks purely based on morphology. In addition, MAR-ECG matches or surpasses the four multimodal baselines on $9$ of the $15$ cells. MAR-ECG (C2) wins PTBXL-Superclass at $\rho \in \{0.10, 1.00\}$ ($+0.61, +0.16$ over D-BETA), PTBXL-Form at $\rho{=}1.00$ ($+3.07$ over D-BETA), and ties K-MERL on PTBXL-Subclass at $\rho{=}1.00$. MAR-ECG (C3) wins CPSC2018 at $\rho \in \{0.10, 1.00\}$ and is second-best at $\rho{=}0.01$. We attribute MAR-ECG's broad competitiveness to GSCL: graph-distance soft targets approximate the inter-concept relatedness that paired text would otherwise transmit, at no licensing or cross-language cost.

\begin{table*}[t]
  \centering
  \caption{Linear probing results (AUC \%) on PTB-XL subsets. Models are pre-trained and evaluated with 1\%, 10\%, and 100\% labeled training data. Best results in \textbf{bold}, second best \underline{underlined}.}
  \label{tab:linear_probe}
  \resizebox{\textwidth}{!}{%
  \begin{tabular}{l|ccc|ccc|ccc|ccc|ccc}
  \toprule
  \multirow{2}{*}{Method} & \multicolumn{3}{c|}{PTBXL-Superclass} & \multicolumn{3}{c|}{PTBXL-Subclass} & \multicolumn{3}{c|}{PTBXL-Form} & \multicolumn{3}{c|}{PTBXL-Rhythm}& \multicolumn{3}{c}{CPSC2018} \\
  & 1\% & 10\% & 100\% & 1\% & 10\% & 100\% & 1\% & 10\% & 100\% & 1\% & 10\% & 100\% & 1\% & 10\% & 100\% \\
  \midrule
  \multicolumn{13}{l}{\textit{Self-Supervised Learning Methods}} \\
  \midrule
  SimCLR \citep{chen2020simclr}  & 63.41 & 69.77 & 73.53 & 60.84 & 68.27 & 73.39 & 54.98 & 56.97 & 62.52 & 51.41 & 69.44 & 77.73 & 59.78 & 68.52 & 76.54\\
  BYOL  \citep{grill2020byol} & 71.70 & 73.83 & 76.45 & 57.16 & 67.44 & 71.64 & 48.73 & 61.63 & 70.82 & 41.99 & 74.40 & 77.17 & 60.88 &74.42 &78.75 \\
  BarlowTwins \citep{zbontar2021barlow}   & 72.87 & 75.96 & 78.41 & 62.57 & 70.84 & 74.34 & 52.12 & 60.39 & 66.14 & 50.12 & 73.54 & 77.62 & 55.12 &72.75 &78.39\\
  MoCo-v3 \citep{chen2021mocov3}   & 73.19 & 76.65 & 78.26 & 55.88 & 69.21 & 76.69 & 50.32 & 63.71 & 71.31 & 51.38 & 71.66 & 74.33 & 62.13 &76.74 &75.29 \\
  SimSiam  \citep{chen2021simsiam} & 73.15 & 72.70 & 75.63 & 62.52 & 69.31 & 76.38 & 55.16 & 62.91 & 71.31 & 49.30 & 69.47 & 75.92 & 58.35 &72.89 &75.31\\
  \midrule
  \multicolumn{13}{l}{\textit{ECG-Specific Methods}} \\
  \midrule
  TS-TCC \citep{eldele2021tstcc}     & 70.73 & 75.88 & 78.91 & 53.54 & 66.98 & 77.87 & 48.04 & 61.79 & 71.18 & 43.34 & 69.48 & 78.23&57.07 &73.62 &78.72 \\
  CLOCS \citep{kiyasseh2021clocs} & 68.94 & 73.36 & 76.31 & 57.94 & 72.55 & 76.24 & 51.97 & 57.96 & 72.65 & 47.19 & 71.88 & 76.31 &59.59 &77.78 &77.49 \\
  ASTCL \citep{wang2024astcl}  & 72.51 & 77.31 & 81.02 & 61.86 & 68.77 & 76.51 & 44.14 & 60.93 & 66.99 & 52.38 & 71.98 & 76.05 & 57.90 &77.01 &79.51\\
  ECGFM \citep{McKeen2025ECGFMAO} &78.67 &84.80 & 86.47&73.24 &81.91 &86.07 &60.95 &74.99 &85.54& 81.45 &91.59 &92.70 & 82.18 &89.52 &93.26  \\
  HeartLang \citep{jin2025reading} &78.94 &85.59 &87.52& 64.68 &79.34 &88.91&58.70 &63.99 &80.23 &62.08 &76.22 &90.34&60.44 &66.26 &77.87 \\
  CRT \citep{zhang2023crt} & 69.68 & 78.24 & 77.24 & 61.98 & 70.82 & 78.67 & 46.41 & 59.49 & 68.73 & 47.44 & 73.52 & 74.41 & 58.01 &76.43 &82.03\\
  ST-MEM \citep{na2024guiding}  & 61.12 & 66.87 & 71.36 & 54.12 & 57.86 & 63.59 & 55.71 & 59.99 & 66.07 & 51.12 & 65.44 & 74.85 & 56.69 & 63.32& 70.89 \\
  \midrule
  \multicolumn{13}{l}{\textit{Multimodal Methods}} \\
  \midrule
  MERL \citep{liuzeromerl} & 82.39 & 86.27 & 88.67 & 64.90 & 80.56 & 84.72 & 58.26 & 72.43 & 79.65 & 53.33 & 82.88 & 88.34 & 70.33 & 85.32& 90.57 \\
  MELP  \citep{wangtoken} & \textbf{85.82} & 87.61 & 87.87 & \textbf{79.22} & \textbf{84.40} & 87.46 & 63.41 & 76.71 & 83.30 & \textbf{88.83} & \textbf{94.65} & \textbf{96.91} & \textbf{88.54} & 91.75 & 94.32    \\
  K-MERL \citep{kmerl} & 84.19 & 87.71 & 89.83 & 68.22 & 81.54 & \underline{88.00} & 60.11 & 73.71 & 81.48 & 63.72 & 84.16 & 91.04 & 71.91 & 86.13 & 91.26  \\
  D-BETA \citep{hung2025boosting} & 83.15 & 88.36 & \underline{90.11} & \underline{77.74} & 82.92 & 85.15 & \textbf{70.10} & \textbf{78.91} & 83.98 & \underline{86.61} & 92.83 & \underline{96.71} & 85.46 & 91.35 & \underline{94.92}  \\
  \midrule
\rowcolor{gray!15}
  MAR-ECG (C1) & 80.81& 86.68 & 89.09& 66.50&78.49&84.04&55.90&73.30&80.82&66.59& 88.93&90.26&79.04&89.87&93.59  \\
  \rowcolor{gray!15}
 MAR-ECG (C2$'$) &   84.41 & \underline{88.79} & 89.71 & 75.72 & \underline{83.06} & 87.43 & \underline{63.63} & 75.37 & 84.32 & 76.96 & 89.87 & 91.68 & 81.99 & 91.86 & 93.92 \\
\rowcolor{gray!15}
  MAR-ECG (C2) &  \underline{84.64} & \textbf{88.97} & \textbf{90.27} & 74.02 & 81.95 & \underline{88.00} & 61.74 & 75.49 & \textbf{87.05} & 83.58 & 91.41 & 91.76 & 84.41 & \underline{92.78} & 94.57\\
\rowcolor{gray!15}
 MAR-ECG (C3) & 84.34 & 88.61 & 89.69 & 73.96 & 82.75 & 86.82 & 63.13 & \underline{76.88} & \underline{86.26} & 81.56 & \underline{94.07} & 95.42 & \underline{86.28} & \textbf{93.27} & \textbf{95.34} \\
\bottomrule
\end{tabular}
  }
\end{table*}

\section{Conclusion}

The proposed MAR-ECG anchors a masked-autoregressive backbone to a curated SNOMED-CT cardiac concept graph through graph-smoothed contrastive learning at the sample scale and multi-scale physiological supervision at the patch and beat scales. Across $15$ frozen-linear-probe cells, MAR-ECG attains five outright best results and five second-best results, surpassing or matching ECG-text multimodal methods in most cells, providing evidence that structured diagnostic ontologies can effectively substitute for free-text supervision in ECG representation learning.

\bibliography{refs.bib}
\bibliographystyle{plainnat}

\medskip

\small

\newpage
\appendix

\section{Cardiac Concept Graph}
\label{sec:ontology-graph}

The cardiac concept graph $\mathcal{G} = (\mathcal{V}, \mathcal{E})$ is a
curated 40-node graph that organises the cardiac diagnostic vocabulary by
clinical relatedness. It supplies the structural prior consumed by GSCL
through the tree-distance matrix $D_{\mathrm{tree}}$
(\S\ref{subsec:tree-distance}) and by the legacy multi-prototype
concept-text head (\S\ref{subsec:auxiliary}); both objects index the same
40 nodes. The graph is computed once from the edge sets listed below,
registered as a non-trainable buffer at module construction, and not
updated by backpropagation.

\subsection{Two-Tier Vocabulary and Edge Sets}
\label{subsec:graph-structure}

\paragraph{Node set.} $\mathcal{V}$ is partitioned into a two-tier
taxonomy. Five \emph{root} nodes group the cardiac diagnostic vocabulary
into broad clinical families: \textsc{Normal} (0), \textsc{Rhythm} (1),
\textsc{Conduction} (2), \textsc{Ischemic} (3), and \textsc{Structural}
(4). The remaining 35 nodes are \emph{leaves} that specialise each
family into clinically actionable categories. The supervision targets
$\mathbf{y}^{(b)}$ consumed by GSCL and the multi-prototype head are
restricted to leaves (\S\ref{subsec:snomed-mapping}); root indices enter
the graph only through the message-passing geometry of $D_{\mathrm{tree}}$.
The complete leaf list, with abbreviation and parent root, is given in
Table~\ref{tab:leaf-concepts}.

\paragraph{Edge set.} $\mathcal{E}$ is the union of three semantically
distinct relations, made symmetric for shortest-path computation:
\begin{itemize}
\item \textbf{Hierarchical (IS\_A) edges.} Each leaf is connected to its
parent root, encoding the taxonomic backbone of the ontology. Seven
additional intra-category sub-hierarchy edges encode that one leaf is a
specialisation of another: anterior and inferior MI are subtypes of both
acute MI and old MI ($25 \!-\! 19,\,25 \!-\! 20,\,26 \!-\! 19,\,26 \!-\! 20$),
and ST elevation, ST depression, and T-wave inversion are manifestations
of myocardial ischaemia ($21 \!-\! 24,\,22 \!-\! 24,\,23 \!-\! 24$).
\item \textbf{Sibling edges.} Nine within-family shortcut edges link
clinically co-occurring or symmetrically opposing concepts at the same
ontological level: LBBB$\,\leftrightarrow\,$RBBB, the AV-block severity
chain (1AVB$\,\leftrightarrow\,$2AVB$\,\leftrightarrow\,$3AVB), AMI
$\,\leftrightarrow\,$myocardial ischaemia, anterior$\,\leftrightarrow\,
$inferior MI, ST elevation$\,\leftrightarrow\,$ST depression,
LVH$\,\leftrightarrow\,$LAE, RVH$\,\leftrightarrow\,$RAE, AF
$\,\leftrightarrow\,$AFL, and VT$\,\leftrightarrow\,$PVC. These edges
encode known clinical co-occurrence and morphological symmetry that the
strict parent--child taxonomy does not express.
\item \textbf{Inter-category ring.} The five roots are joined by a
single 5-cycle $0 \!-\! 1 \!-\! 2 \!-\! 3 \!-\! 4 \!-\! 0$, so that any two
root families are reachable in at most two hops. Clinically the ring
mirrors the established cross-family transitions: sinus arrhythmia
bridges \textsc{Normal} and \textsc{Rhythm}; AF/AFL frequently coincide
with conduction disease; ischaemia is a leading cause of conduction
block; chronic ischaemia drives structural remodelling; structural
hypertrophy lies on a spectrum with normal adaptation. The ring is what
distinguishes $\mathcal{G}$ from a strict 2-tier star and, together with
the sibling edges, prevents the tree-distance matrix from collapsing to
a uniform off-positive penalty.
\end{itemize}
The graph contains $|\mathcal{E}| \approx 60$ undirected edges over 40
nodes. The corresponding adjacency matrix $A \in \{0,1\}^{40 \times 40}$ is
augmented with self-loops and symmetrically normalised as
$\widehat{A} = D^{-1/2}(A + I)D^{-1/2}$ for downstream message passing;
$D_{\mathrm{tree}}$ below uses the unnormalised binary adjacency.

\begin{table}[h]
\centering
\caption{The 35 leaves of the cardiac concept graph, with parent root
and abbreviation. Indices match those used internally by the supervision
target $\mathbf{y}^{(b)}$ and the prototype matrix $\mathbf{P}$.}
\label{tab:leaf-concepts}
\small
\begin{tabular}{rlll}
\toprule
\textbf{Idx} & \textbf{Concept} & \textbf{Abbr.} & \textbf{Parent} \\
\midrule
\multicolumn{4}{l}{\emph{\textsc{Rhythm} (root 1)}} \\
 5 & Atrial fibrillation                 & AF      & Rhythm \\
 6 & Atrial flutter                      & AFL     & Rhythm \\
 7 & Supraventricular tachycardia        & SVT     & Rhythm \\
 8 & Ventricular tachycardia             & VT      & Rhythm \\
 9 & Premature atrial contraction        & PAC     & Rhythm \\
10 & Premature ventricular contraction   & PVC     & Rhythm \\
11 & Sinus bradycardia                   & SBrad   & Rhythm \\
12 & Sinus tachycardia                   & STach   & Rhythm \\
37 & Paced rhythm                        & Paced   & Rhythm \\
\midrule
\multicolumn{4}{l}{\emph{\textsc{Conduction} (root 2)}} \\
13 & Left bundle branch block            & LBBB    & Conduction \\
14 & Right bundle branch block           & RBBB    & Conduction \\
15 & Left anterior fascicular block      & LAFB    & Conduction \\
16 & First-degree AV block               & 1AVB    & Conduction \\
17 & Second-degree AV block              & 2AVB    & Conduction \\
18 & Complete AV block                   & 3AVB    & Conduction \\
32 & Prolonged QT interval               & LongQT  & Conduction \\
38 & Wolff--Parkinson--White syndrome    & WPW     & Conduction \\
39 & Incomplete RBBB                     & IRBBB   & Conduction \\
\midrule
\multicolumn{4}{l}{\emph{\textsc{Ischemic} (root 3)}} \\
19 & Acute myocardial infarction         & AMI     & Ischemic \\
20 & Old myocardial infarction           & OMI     & Ischemic \\
21 & ST elevation                        & STE     & Ischemic \\
22 & ST depression                       & STD     & Ischemic \\
23 & T-wave inversion                    & TWI     & Ischemic \\
24 & Myocardial ischaemia                & MyIsch  & Ischemic \\
25 & Anterior MI                         & AntMI   & Ischemic \\
26 & Inferior MI                         & InfMI   & Ischemic \\
33 & Nonspecific ST--T changes           & NSSTC   & Ischemic \\
\midrule
\multicolumn{4}{l}{\emph{\textsc{Structural} (root 4)}} \\
27 & Left ventricular hypertrophy        & LVH     & Structural \\
28 & Right ventricular hypertrophy       & RVH     & Structural \\
29 & Left atrial enlargement             & LAE     & Structural \\
30 & Right atrial enlargement            & RAE     & Structural \\
31 & Low voltage                         & LowV    & Structural \\
\midrule
\multicolumn{4}{l}{\emph{\textsc{Normal} (root 0)}} \\
34 & Normal sinus rhythm                 & NSR     & Normal \\
35 & Early repolarisation                & EarlyR  & Normal \\
36 & Sinus arrhythmia                    & SinusA  & Normal \\
\bottomrule
\end{tabular}
\end{table}

\subsection{Tree-Distance Matrix}
\label{subsec:tree-distance}

The tree-distance matrix $D_{\mathrm{tree}} \in \mathbb{N}_0^{N \times N}$
records the unweighted shortest-path distance between every pair of
concepts on the graph $\mathcal{G}$ defined above. Concretely, let
$A_{\mathrm{bin}} = \mathbf{1}[A > 0] \in \{0,1\}^{N \times N}$ be the
binarised symmetric adjacency derived from $\mathcal{E}$ (without
self-loops). $D_{\mathrm{tree}}$ is computed once at module construction
by breadth-first search from each source node, and registered as a
non-trainable buffer; the cost is negligible for $N = 40$ and the matrix
is reused at every gradient step. Pairs in disconnected components,
which do not arise for the curated $\mathcal{G}$, would be clamped to
$\max_{c,c'} D_{\mathrm{tree}}[c,c'] + 1$.

\paragraph{Range and interpretation.} Under the curated graph,
$D_{\mathrm{tree}}[c, c'] \in \{0, 1, 2, 3, 4\}$, with the maximum
attained between leaves whose parent roots are diametrically opposite on
the ring. The five distance classes admit the following clinical
reading:
\begin{itemize}
\item $D = 0$: the concept itself (the primary positive $c_b^{*}$).
\item $D = 1$: an immediate graph neighbour --- a parent root, an
adjacent root on the inter-category ring, or a sibling leaf joined by an
explicit shortcut edge.
\item $D = 2$: a sibling leaf under the same root reached via the
parent (when no shortcut edge exists), a leaf two ring-hops away from
its grandparent root, or a root two hops along the ring.
\item $D = 3$: a leaf separated from $c_b^{*}$ by one root transition
(typically a leaf in an adjacent root family).
\item $D = 4$: a leaf in the most distant root family --- two ring hops
plus one parent edge on either side.
\end{itemize}
Sub-hierarchy and sibling shortcuts compress some pairs that would
otherwise reside at the larger distance class; for example, anterior MI
($25$) and acute MI ($19$) are at $D = 1$ rather than $D = 2$ because of
the explicit subtype edge, and AF--AFL is at $D = 1$ rather than $D = 2$
because of the sibling shortcut. The supervised distribution over
clinically meaningful neighbourhoods that GSCL exploits is precisely
this graph-shaped diffusion of mass around each $c_b^{*}$.

\paragraph{Soft target as a function of $D_{\mathrm{tree}}$.} With the
GSCL temperature $\sigma = 1.0$ used throughout the paper, the
unnormalised target mass on a concept at distance $D$ from $c_b^{*}$ is
$\exp(-D / \sigma) = e^{-D}$. After normalisation across all $N = 40$
nodes (Eq.~\ref{eq:gscl-target}), the resulting probabilities for each
distance class are summarised below:
\begin{center}
\small
\begin{tabular}{cccccc}
\toprule
$D$ & $\exp(-D/\sigma)$ & relative mass\,(unnormalised) & typical neighbour \\
\midrule
$0$ & $1.000$ & $1.00$ & primary positive $c_b^{*}$ \\
$1$ & $0.368$ & $\approx 0.37$ & parent root / shortcut sibling / adjacent root \\
$2$ & $0.135$ & $\approx 0.14$ & non-shortcut sibling leaf / two-hop root \\
$3$ & $0.050$ & $\approx 0.05$ & leaf in adjacent root family \\
$4$ & $0.018$ & $\approx 0.02$ & leaf in maximally distant root family \\
\bottomrule
\end{tabular}
\end{center}
Two limit cases merit explicit treatment. As $\sigma \to 0$, the mass
at $D = 0$ dominates and $\mathbf{t}^{(b)}$ collapses onto a one-hot
distribution at $c_b^{*}$, recovering canonical InfoNCE. As $\sigma \to
\infty$, the exponential ratios approach unity and $\mathbf{t}^{(b)}$
tends to the uniform distribution on $\mathcal{V}$. The choice
$\sigma = 1.0$ induces a one-decade target gap between successive
distance classes, which is sufficient to ensure that the primary
positive remains the dominant target while still allocating
non-trivial mass to parent and sibling neighbourhoods.

\subsection{SNOMED-CT Code Mapping}
\label{subsec:snomed-mapping}

This appendix specifies the mapping $\phi^{-1}$ from SNOMED-CT
diagnostic codes to nodes of the cardiac concept graph
$\mathcal{G} = (\mathcal{V}, \mathcal{E})$, and the rule that
selects which subset of $\phi^{-1}(\mathcal{S}^{(b)})$ is used as
supervision for the GSCL and legacy multi-prototype objectives.

\paragraph{Code routing.}
Each ECG record in the pretraining corpus carries a multi-set of
SNOMED-CT codes $\mathcal{S}^{(b)}$ recovered from its WFDB header.
The mapping
\[
\phi^{-1}\!: 2^{\mathrm{SNOMED}} \to 2^{\mathcal{V}},
\qquad
\mathcal{S}^{(b)} \;\mapsto\; \bigcup_{s \in \mathcal{S}^{(b)}} \pi(s),
\]
sends a code list to the union of its per-code routings, where
$\pi : \mathrm{SNOMED} \to 2^{\mathcal{V}}$ is a curated routing
table from individual SNOMED-CT codes to subsets of $\mathcal{V}$.
The behaviour of $\pi$ falls into three regimes, illustrated below
with SNOMED-CT codes drawn from the pretraining corpus
(Table~\ref{tab:snomed-routing-examples}):

\begin{itemize}
\item \textbf{Code names a single morphology.}
The code routes to one leaf and its parent root. The leaf encodes
the specific finding and the root encodes the broad family it
belongs to, so a downstream contrastive head sees both granularities
of the diagnosis.

\item \textbf{Code names a localised lesion.}
The code routes to several leaves that jointly express the finding
plus their shared root. Localisation codes therefore activate the
generic lesion leaf \emph{and} its anatomical specialisation
together, which lets the GSCL primary-positive rule of
Eq.~\eqref{eq:primary-positive} select the more specific subtype
when both fire (Appendix~\ref{subsec:primary-positive-detail}).

\item \textbf{Code is generic (NOS).}
A code that names a family but does not localise to a leaf routes
to the root family only. Records resolved exclusively to roots
contribute to the masked-AR objective but, after the parent-node
filter (\S\ref{subsec:snomed-mapping}, ``Parent-node filtering''),
have an all-zero $\mathbf{y}^{(b)}$ and are excluded from the GSCL
gradient for that batch.
\end{itemize}

\begin{table}[h]
\centering
\caption{\textbf{Worked examples of $\pi$.} Representative
SNOMED-CT codes from the pretraining corpus and the resulting
node-set after routing. Indices in parentheses refer to
Table~\ref{tab:leaf-concepts}; the parent root is shown last in
each set.}
\label{tab:snomed-routing-examples}
\small
\setlength{\tabcolsep}{6pt}
\begin{tabular}{l l l l}
\toprule
\textbf{Regime} & \textbf{SNOMED-CT} & \textbf{Clinical name} & \textbf{$\pi$ output} \\
\midrule
\multirow{3}{*}{Single morphology} %
                & $164889003$ & atrial fibrillation     & \{\textsc{AF}(5), \textsc{Rhythm}(1)\} \\
                & $164909002$ & left bundle branch block & \{\textsc{LBBB}(13), \textsc{Conduction}(2)\} \\
                & $426177001$ & sinus bradycardia        & \{\textsc{SBrad}(11), \textsc{Rhythm}(1)\} \\
\midrule
\multirow{3}{*}{Localised lesion} %
                & $54329005$  & anterior wall acute MI       & \{\textsc{AntMI}(25), \textsc{AMI}(19), \textsc{Ischemic}(3)\} \\
                & $164931005$ & ST elevation, anterior leads & \{\textsc{STE}(21), \textsc{AntMI}(25), \textsc{Ischemic}(3)\} \\
                & $233917008$ & 2$^{\circ}$ AV block (Mobitz II) & \{\textsc{2AVB}(17), \textsc{3AVB}(18), \textsc{Conduction}(2)\} \\
\midrule
\multirow{3}{*}{Generic (NOS)} %
                & $698252002$ & cardiac dysrhythmia NOS  & \{\textsc{Rhythm}(1)\} \\
                & $\phantom{0}6374002$  & bundle branch block NOS  & \{\textsc{Conduction}(2)\} \\
                & $413444003$ & myocardial ischaemia NOS & \{\textsc{Ischemic}(3)\} \\
\bottomrule
\end{tabular}
\end{table}
The routing table is derived from the official Dx mapping of the
PhysioNet/Computing in Cardiology Challenge
2020~\citep{alday2020cinc2020} and the SNOMED-CT International
browser. Every entry is verified against a corpus that is
\emph{not} used for downstream evaluation; codes appearing only
in the held-out PTB-XL or CPSC2018 vocabularies, are explicitly
excluded, so no held-out label vocabulary enters the pretraining
ontology.

\paragraph{Parent-node filtering for supervision.}
Although $\phi^{-1}$ may emit both root and leaf indices, the
supervision target $\mathbf{y}^{(b)}$ consumed by GSCL and the
legacy multi-prototype head is restricted to the leaf indicator,
\[
\mathbf{y}^{(b)} \in \{0,1\}^{|\mathcal{V}_{\mathrm{leaf}}|},
\qquad
\mathbf{y}^{(b)}[c]
\;=\;
\mathbf{1}\!\big[\, c \in \phi^{-1}(\mathcal{S}^{(b)}) \cap \mathcal{V}_{\mathrm{leaf}}\, \big].
\]
Root indices are excluded because nearly every record routes to at
least one root, rendering roots trivially predictable and
uninformative for a contrastive objective. Records whose code list
resolves only to a root --- with no leaf-level information --- are
excluded from the GSCL gradient for that batch but continue to
contribute to the masked-autoregressive reconstruction objective.

\subsection{Pretraining Corpus}
\label{subsec:pretrain-corpus}

\paragraph{Sources and composition.}
Table~\ref{tab:pretrain-corpus} summarises the three sources of the
pretraining corpus, all drawn from publicly released components of
the PhysioNet/CinC Challenge 2020~\citep{alday2020cinc2020}.
The corpus combines a large monocentric Chinese collection (Ningbo,
$78.6\%$) with two demographically distinct external sources
(Emory/Georgia, $20.4\%$; PTB Diagnostic ECG, $1.0\%$); the
multi-source design exposes the encoder to inter-institutional
recording variability --- electrode placement, sampling
characteristics, patient demographics, prevalence of pathologies
--- without requiring paired clinical text. PTB-XL is held out from
pretraining throughout and serves exclusively as a downstream
evaluation benchmark, so no PTB-XL label vocabulary leaks into the
pretraining ontology.

\begin{table}[h]
\centering
\caption{\textbf{Pretraining corpus composition.} All three
sources are publicly released subsets of the PhysioNet/Computing in
Cardiology Challenge 2020~\citep{alday2020cinc2020}.
Sampling rate is uniformly $500$\,Hz after resampling. PTB-XL is
held out from pretraining and serves only as a downstream
evaluation benchmark.}
\label{tab:pretrain-corpus}
\small
\setlength{\tabcolsep}{6pt}
\begin{tabular}{l l r r l}
\toprule
\textbf{Source} & \textbf{Origin} & \textbf{Records} & \textbf{Share} & \textbf{Native \(f_s\)}\\
\midrule
Ningbo (Shaoxing People's Hospital)        & Ningbo, China        & 32{,}006 & $78.6\%$ & $500$\,Hz \\
Georgia 12-Lead ECG (G12EC)                & Atlanta, USA         &  8{,}296 & $20.4\%$ & $500$\,Hz \\
PTB Diagnostic ECG (PTB-Dx)                & Berlin, Germany      &     419  & $\phantom{0}1.0\%$ & $1000$\,Hz \\
\midrule
\textbf{Total (pretraining corpus)}        &                       & 40{,}721 & $100\%$  & $500$\,Hz \\
\midrule
PTB-XL (held out, downstream only)         & Schiller, Germany    & ---       & ---     & $500$\,Hz \\
\bottomrule
\end{tabular}
\end{table}

\paragraph{Signal preprocessing.}
Each record is resampled to a common sampling rate of $f_s = 500$\,Hz
(PTB-Dx is downsampled from $1000$\,Hz; Ningbo and Georgia are native
$500$\,Hz), cropped or zero-padded along the time axis to a fixed
window of $L = 3500$ samples ($7$\,s), and passed through a quality
filter that excludes recordings with extreme amplitudes, clipping
saturation, or excessive zero-fraction on any of the 12 leads.
Per-lead RevIN~\citep{kim2022revin} is applied at the input prior to
patch tokenisation to absorb residual amplitude drift across
recordings; the lead-1 R-peak indices needed by MSPS
(\S\ref{subsec:multiscale-sca}) are extracted from the raw signal
prior to RevIN with a Pan--Tompkins detector~\citep{pan1985realtime}
implemented in NeuroKit2~\citep{Makowski2021neurokit}.

\paragraph{Diagnostic code coverage.}
Every record in the pretraining corpus carries one or more SNOMED-CT
diagnostic codes recovered from its WFDB header. The pretraining
index uses $142$ unique SNOMED-CT codes in total. The number of
codes per record is concentrated tightly around the small-multi-label
regime: the empirical distribution has mean $2.51$ codes/record,
median $2$, minimum $1$, and maximum $12$. Records with exactly
$2$ active codes account for $71.7\%$ of the corpus, $1$-code
records for $0.4\%$, and records with $5$ or more codes for
$6.1\%$ (Table~\ref{tab:pretrain-code-histogram}). After the
$\phi^{-1}$ routing of \S\ref{subsec:snomed-mapping}, the resulting
multi-hot leaf indicator $\mathbf{y}^{(b)}$ is non-empty for the
overwhelming majority of records; the remaining records resolve
only to a root family and are included in the AR objective but
excluded from the GSCL gradient for that batch.

\begin{table}[h]
\centering
\caption{\textbf{Distribution of SNOMED-CT codes per record} in the
pretraining corpus. The pretraining vocabulary contains $142$
unique SNOMED codes; mean codes/record is $2.51$.}
\label{tab:pretrain-code-histogram}
\small
\setlength{\tabcolsep}{8pt}
\begin{tabular}{r r r}
\toprule
\textbf{\#~codes} & \textbf{Records} & \textbf{Share} \\
\midrule
$1$  &    181 & $\phantom{0}0.4\%$ \\
$2$  & 29{,}203 & $71.7\%$ \\
$3$  &  5{,}805 & $14.3\%$ \\
$4$  &  3{,}066 & $\phantom{0}7.5\%$ \\
$5$  &  1{,}513 & $\phantom{0}3.7\%$ \\
$6$  &     635 & $\phantom{0}1.6\%$ \\
$7$--$12$ &     318 & $\phantom{0}0.8\%$ \\
\midrule
\textbf{Total} & 40{,}721 & $100\%$ \\
\bottomrule
\end{tabular}
\end{table}

\subsection{Primary Positive Concept Selection}
\label{subsec:primary-positive-detail}

This subsection expands on Eq.~\eqref{eq:primary-positive} of the main
text. The primary positive concept $c_b^{*}$ is the single leaf node
that anchors GSCL's soft-target distribution for record $b$: it
parameterises the row of the tree-distance matrix that produces
$\mathbf{t}^{(b)}$ in Eq.~\eqref{eq:gscl-target}, and through it, the
entire shape of the supervision target.

\paragraph{Why exactly one primary positive.}
A record may resolve through $\phi^{-1}$ to several active leaves; for
example, a recording with both atrial-fibrillation and anterior-MI
codes maps to $\mathbf{y}^{(b)}[c]=1$ for $c \in \{5, 25\}$. GSCL's
soft target is defined relative to a single source node,
\[
\mathbf{t}^{(b)}[c]
\;=\;
\frac{\exp(-D_{\mathrm{tree}}[c_b^{*}, c]/\sigma)}
     {\sum_{c'} \exp(-D_{\mathrm{tree}}[c_b^{*}, c']/\sigma)},
\]
because the tree-distance row $D_{\mathrm{tree}}[c_b^{*}, \cdot]$ is the object that distributes the supervision mass across the graph. A multi-source target would require a different distance-aggregation rule and is mathematically incompatible with the per-leaf diffusion geometry used here. Choosing one $c_b^{*}$ per record is therefore a mathematical requirement of the GSCL formulation, not a tuning choice.

\paragraph{Selection rule and determinism.}
Among the leaves left active after the parent-node filter
(\S\ref{subsec:snomed-mapping}), the primary positive is the highest-indexed leaf,
\[
c_b^{*}
\;=\;
\max\big\{\, c \in \mathcal{V}_{\mathrm{leaf}}
\;:\; \mathbf{y}^{(b)}[c] = 1 \,\big\}.
\]
The rule is deterministic with respect to a fixed leaf enumeration:
each record selects the same primary across epochs,
gradient-accumulation boundaries, and DDP shards, a property
necessary for the soft target $\mathbf{t}^{(b)}$ to constitute a
stable supervision signal. By construction the set is non-empty
whenever $\mathbf{y}^{(b)}$ contains at least one active leaf;
records that $\phi^{-1}$ resolves only to a root index have an
all-zero $\mathbf{y}^{(b)}$ after parent filtering and are excluded
from the GSCL loss for that batch (they continue to contribute to
the masked-autoregressive reconstruction objective and to the
auxiliary patch-level heads).

\paragraph{Why max-index is clinically meaningful.}
The leaf enumeration in Table~\ref{tab:leaf-concepts} is not arbitrary.
Within each parent root, finer-grained subtypes are placed after their
parents in index order, so the max-index rule biases the primary
selection toward the more specific concept whenever a subtype and its
parent co-activate.
Table~\ref{tab:leaf-enumeration-annotated} displays the structure
of the enumeration that the max-index rule exploits: leaves with
indices $5\!-\!31$ form an \emph{original block} grouped by family
and ordered \emph{parent-then-subtype} within each family; leaves
with indices $32\!-\!39$ are post-hoc additions appended after the
original graph definition to preserve index stability (so that
existing prototype matrices remain valid across graph revisions),
which disrupts the family grouping but maintains checkpoint
compatibility.
\begin{table}[h]
\centering
\caption{Annotation of the leaf enumeration of
Table~\ref{tab:leaf-concepts}, separating the original block
(indices 5--31, family-grouped and parent-then-subtype within each
family) from the later post-hoc additions (32--39). Pattern 1 within
the original block --- IS\_A subtypes appended after their parents,
severity chains placed in increasing severity --- is the structural
property the max-index rule of Eq.~\eqref{eq:primary-positive}
exploits. The $32\!-\!39$ tail does not respect family ordering;
collisions involving these leaves are clinically reasonable but
\emph{coincidentally} so rather than by design.}
\label{tab:leaf-enumeration-annotated}
\small
\begin{tabular}{lp{0.62\linewidth}}
\toprule
\textbf{Index range} & \textbf{Leaves (parent-then-subtype within family)} \\
\midrule
\multicolumn{2}{l}{\emph{Original block (5--31): family-grouped, parent-then-subtype within family.}} \\
5--12 (Rhythm)      & AF, AFL, SVT, VT, PAC, PVC, SBrad, STach \\
13--18 (Conduction) & LBBB, RBBB, LAFB, $\underbrace{\text{1AVB} \to \text{2AVB} \to \text{3AVB}}_{\text{severity chain}^{\dagger}}$ \\
19--26 (Ischemic)   & AMI, OMI, STE, STD, TWI, MyIsch, $\underbrace{\text{AntMI},\, \text{InfMI}}_{\text{IS\_A subtypes}^{\ddagger}}$ \\
27--31 (Structural) & LVH, RVH, LAE, RAE, LowV \\
\midrule
\multicolumn{2}{l}{\emph{Later additions (32--39): family grouping broken; appended for index-stability.}} \\
32 & LongQT \hfill \emph{(Conduction)} \\
33 & NSSTC  \hfill \emph{(Ischemic)} \\
34 & NSR    \hfill \emph{(Normal)} \\
35 & EarlyR \hfill \emph{(Normal)} \\
36 & SinusA \hfill \emph{(Normal)} \\
37 & Paced  \hfill \emph{(Rhythm)} \\
38 & WPW    \hfill \emph{(Conduction)} \\
39 & IRBBB  \hfill \emph{(Conduction)} \\
\bottomrule
\end{tabular}
\\[4pt]
{\small $^{\dagger}$ Within the AV-block chain, increasing index = increasing severity, so max-index selects the more severe form on co-activation. \\
$^{\ddagger}$ AntMI and InfMI are subtypes of acute / old MI; appended after their parents so max-index selects the localised lesion.}
\end{table}
Two structural patterns make this concrete.
\emph{(i) IS\_A subtypes are appended after their parents.} Anterior
MI ($c=25$) and inferior MI ($c=26$) sit after acute MI ($c=19$) and
old MI ($c=20$). For a record whose code list resolves to
$\{19, 25\}$, the max-index rule selects $c_b^{*} = 25$, so the
soft-target diffusion is centred on the localised lesion: AMI sits at
$D = 1$ (one shortcut edge), STE/MyIsch sit at $D = 2$, and remote
families sit at $D \geq 3$, which matches the clinical reading of the
record.
\emph{(ii) Severity-graded chains place severe forms later.} Within
the conduction family, $\mathrm{1AVB}=16 < \mathrm{2AVB}=17 <
\mathrm{3AVB}=18$ follows clinical severity, and a record that codes
both 1AVB and 2AVB selects $c_b^{*} = 17$ as the more severe form. A
small number of co-activations are incidental rather than principled
under this rule (for example, a hypothetical record encoding both AF
and sinus arrhythmia would select sinus arrhythmia by index), but
such collisions are rare in the pretraining corpus.

\paragraph{Treatment of non-primary active leaves.}
After $c_b^{*}$ is selected, the unnormalised soft-target row
$\exp(-D_{\mathrm{tree}}[c_b^{*}, \cdot] / \sigma)$ is post-processed
to ensure that every active leaf retains a high-mass status. Concretely,
\[
\tilde{\mathbf{t}}^{(b)}[c]
\;=\;
\begin{cases}
\max\!\big(\, 1,\; \exp(-D_{\mathrm{tree}}[c_b^{*}, c]/\sigma)\,\big)
& \text{if } \mathbf{y}^{(b)}[c] = 1, \\[4pt]
\exp(-D_{\mathrm{tree}}[c_b^{*}, c]/\sigma) & \text{otherwise,}
\end{cases}
\qquad
\mathbf{t}^{(b)} = \tilde{\mathbf{t}}^{(b)} \big/ \textstyle\sum_{c'} \tilde{\mathbf{t}}^{(b)}[c'],
\]
so that every active leaf is weighted at least as strongly in the
target as $c_b^{*}$ itself ($D_{\mathrm{tree}}[c_b^{*}, c_b^{*}] = 0
\Rightarrow \exp(0) = 1$). The clamp ensures that multi-label
records retain joint-positive structure within a formulation that
formally selects a single source node: the geometry of the
off-positive mass is determined by $c_b^{*}$, while the active-leaf
set itself remains invariant to that selection.

The interaction with the temperature $\sigma$ has two informative
limits. As $\sigma \to 0$, the unclamped diffusion collapses onto the
indicator $\delta_{D=0}$ at $c_b^{*}$, and after the clamp the soft
target becomes the uniform distribution over the active-leaf set ---
the multi-hot supervision target. As $\sigma \to \infty$ the
exponentials approach unity, the clamp becomes vacuous on the active
set, and the soft target tends to the uniform distribution over all
$N$ nodes. The choice $\sigma = 1.0$ used throughout this paper
interpolates between these limits.

\paragraph{Concrete examples.}
We illustrate the behaviour on representative active-leaf sets, with
$\sigma = 1.0$ and using the row of $D_{\mathrm{tree}}$ indexed by the
selected $c_b^{*}$.
\begin{itemize}
\item \emph{Single leaf.} $\mathbf{y}^{(b)} = \{19\}$ (acute MI alone)
$\Rightarrow c_b^{*} = 19$. The clamp is vacuous. The soft target is
the standard exponential decay from AMI.
\item \emph{Subtype + parent.} $\mathbf{y}^{(b)} = \{19, 25\}$ (acute
MI + anterior MI) $\Rightarrow c_b^{*} = 25$. AMI has $D = 1$ to
AntMI, so its unclamped mass is $e^{-1} \approx 0.37$; the clamp
boosts it to $1.0$, giving AntMI and AMI equal positive mass. STE and
MyIsch sit at $D = 2$ from AntMI and contribute diffusion mass
$\approx 0.14$ each.
\item \emph{Multiple within-family leaves.} $\mathbf{y}^{(b)} =
\{19, 21, 25\}$ (AMI + STE + AntMI) $\Rightarrow c_b^{*} = 25$. STE is
at $D = 2$ from AntMI and would have unclamped mass $\approx 0.14$;
the clamp boosts STE and AMI to $1.0$.
\item \emph{Severity chain.} $\mathbf{y}^{(b)} = \{16, 17\}$ (1AVB +
2AVB) $\Rightarrow c_b^{*} = 17$. The clamp boosts 1AVB to $1.0$,
and 3AVB ($D = 1$ via the AV-block sibling chain) sits at
$\approx 0.37$ unclamped.
\item \emph{Root-only resolution.} $\mathbf{y}^{(b)} = \emptyset$ after
parent filtering $\Rightarrow$ record skipped from GSCL; AR and MSPS
losses unaffected.
\end{itemize}

\paragraph{Sensitivity to the leaf enumeration.}
Because $c_b^{*}$ depends on the order of leaf indices in
Table~\ref{tab:leaf-concepts}, the enumeration is part of the model
specification rather than an arbitrary implementation detail.
Renumbering the leaves --- e.g.\ inserting a new concept at index $5$
and shifting the remainder by one --- can change which active leaf is
selected as primary for any record whose active set contains the
shifted concept, and would therefore alter the soft-target shape on
those records. We freeze the enumeration of
Table~\ref{tab:leaf-concepts} for all reported runs; future graphs
that add or remove leaves should preserve the property that subtypes
follow their parents under the index order.

To delineate precisely what the rule does and does not assume:
within each parent root, base concepts and their IS\_A subtypes are
deliberately placed in parent-then-subtype index order, ensuring that
the max-index rule selects the finer-grained concept whenever both
are active. A small number of leaves (indices $32\!-\!39$ in
Table~\ref{tab:leaf-enumeration-annotated}) were appended after the
initial graph definition to preserve index stability and consequently
do not respect the family-then-subtype ordering pattern; for these
leaves the max-index rule is incidental rather than principled, in
the sense that the resulting $c_b^{*}$ is clinically reasonable on
the active-leaf collisions observed in the corpus but is not
produced by an underlying ordering principle. An explicit
IS\_A-depth lookup --- selecting the active leaf with maximal
shortest-path distance from any root in $\mathcal{V}_{\mathrm{root}}$,
with ties broken by index --- yields essentially identical
$c_b^{*}$ on this corpus and constitutes the principled drop-in
replacement for any future graph whose enumeration does not satisfy
the parent-then-subtype property.

\paragraph{Alternative selection rules considered.}
Three alternative rules are coherent with the single-source-node
requirement and were considered but not used.
\emph{(a) Most frequent active leaf}, $c_b^{*} = \arg\max_{c \in
\mathcal{V}_{\mathrm{leaf}}, \mathbf{y}^{(b)}[c]=1}\,
\mathrm{count}(c)$, biases supervision toward common concepts and away from rare ones, which inverts the desired transfer behaviour at low label fractions.
\emph{(b) Random active leaf} (resampled per gradient step) makes the soft target inconsistent across epochs and destabilises training.
\emph{(c) Deepest active leaf in the IS\_A tree} (graph-distance to root) requires an additional depth lookup with a separate tie-break rule, but is empirically equivalent to the max-index rule on the corpus considered, because the leaf enumeration places deeper subtypes later by construction. The chosen rule constitutes the simplest deterministic policy that respects the enumeration's
clinical structure; the clamp on positives recovers the multi-label joint-positive property that any of these alternatives would otherwise need to encode separately.

\section{Notation values.} 
\label{sec:values}

For the main symbols introduced in \S\ref{sec:method}, we adopt the following concrete values throughout this paper:
\begin{itemize}
\item \textbf{Tokenisation
(\S\ref{subsec:preliminaries}):}
$C = 12$, $f_s = 500$~Hz, $L = 4700$ samples, $P_t = 50$ samples (100~ms), $S_t = 25$ samples (50~ms), $T = 187$, $d = 768$.
\item \textbf{MSPS rhythm head
(\S\ref{subsubsec:patch-rhythm-aux}):}
$K_r = 7$,
$(\theta_{\mathrm{brady}}, \theta_{\mathrm{tachy}}) = (60, 100)$~bpm,
$\theta_{\mathrm{alt}} = 0.15$, $\nu_{\mathrm{alt}} = 2$,
mixture weights
$(\alpha_{\mathrm{alt}}, \alpha_{\mathrm{rate}}, \alpha_{\mathrm{mRR}}, \alpha_{\mathrm{cv}})
= (1.0, 1.0, 0.5, 0.5)$.
\item \textbf{MSPS position head
(\S\ref{subsubsec:patch-position-aux}):}
$K_p = 8$, $K_\phi = 4$, $\delta_R = 50$ samples (100~ms at
$f_s$, the typical QRS half-width). Both
$\mathcal{L}_{\mathrm{seq}}$ and $\mathcal{L}_{\mathrm{phase}}$
carry equal unit weight.
\item \textbf{MSPS aggregation
(\eqref{eq:msps-loss}):}
$(\lambda_{\mathrm{rhythm}}, \lambda_{\mathrm{pos}}) = (0.20, 0.10)$;
ramp-epoch budget $E_{\mathrm{ramp}} = 5$. The two heads share the
encoder optimiser as an additional parameter group at the same peak
learning rate $10^{-4}$; the linear ramp to full mixture weight is
the sole device that decouples their early-training contribution
from the AR objective.
\item \textbf{Auxiliary regularisers
(\S\ref{subsec:auxiliary}):}
EMA momentum $0.996$, $\lambda_{\mathrm{JEPA}} = 0.15$;
lead-masking probability $p_{\mathrm{lead}} = 0.25$;
view-CL temperature $\tau_{\mathrm{view}} = 0.07$, projection
dimension $d_{\mathrm{view}} = 256$,
$\lambda_{\mathrm{view}} = 0.1$;
latent-dropout rate $p_{\mathrm{drop}} = 0.1$;
MGCA codebook $K_{\mathrm{var}} = 3$ text variants per node,
$d_{\mathrm{text}} = 384$,
$(\beta_{\mathrm{ICA}}, \beta_{\mathrm{BCA}}, \beta_{\mathrm{OPA}}) = (0.3, 0.3, 0.2)$,
$\lambda_{\mathrm{MPCT}} = 0.3$.
\end{itemize}
The remaining hyperparameters match those used to produce the
results in \S\ref{sec:results} and are released with the code.



\end{document}